\documentclass[runningheads]{svmult}
\usepackage{makeidx}   
\usepackage{graphicx}  
\usepackage{subeqnar}  
\usepackage{multicol}  
\usepackage{cropmark} 
\usepackage{physprbb}  
\begin{document}
\title*{Status of the HDMS experiment, the GENIUS project and the GENIUS-TF}
\toctitle{Status of the HDMS experiment, the GENIUS project and the GENIUS-TF}
%
%
\titlerunning{The HDMS experiment, GENIUS and the GENIUS-TF}
%
\author{H. V. Klapdor-Kleingrothaus$^{\dagger}$
\and B. Majorovits\thanks{talk presented by B. Majorovits}
\and L. Baudis
\and A. Dietz
\and G.~Heusser
\and I. Krivosheina
\and H. Strecker}
\authorrunning{H.V. Klapdor-Kleingrothaus et al.}
%
%
\institute{Max Planck Institut f\"ur Kernphysik, P.O. Box 103980,
  69029 Heidelberg, Germany\\
  \vspace*{0.4cm}
  $^{\dagger}$ Spokesman of the GENIUS Collaboration}
\maketitle              

\begin{abstract}
The status of dark matter search in Heidelberg is reviewed. After one
year of running the HDMS prototype experiment in the Gran Sasso 
Underground Laboratory, the inner crystal of the detector has been
replaced with a HPGe crystal of enriched $^{73}$Ge.  The results of
the operation of the HDMS prototype detector are discussed.
In the light of the contradictive results from the CDMS and DAMA experiments 
the GENIUS-TF, a new experimental setup is proposed. 
The GENIUS-TF could probe the DAMA evidence region using the WIMP
nucleus recoil signal and WIMP annual modulation signature
simulataneously. Besides that it can prove some key parameters 
of the detector technique, to be implemented into the GENIUS setup and
will in this sense be a first step towards the realization of the
GENIUS experiment.
\end{abstract}

\section{Introduction}
The topic of Dark Matter search has lately gained some actuality by the results of the
DAMA \cite{damahere} and CDMS \cite{cdmshere} experiments. 
The DAMA collaboration claims to see positive evidence for WIMP
dark matter using the annual modulation signature, whereas the CDMS experiment almost
fully excludes the DAMA allowed cross-sections
for WIMP dark matter (the 3 $\sigma$ region with 84\% C.L.).    
It is therefore of utmost importance to independently test these results using both
experimental approaches: to look for the WIMP-nucleus recoil signal and for the annual
modulation effect. However, should the positive DAMA WIMP evidence be disproven, a large
step forward in terms of increasing the sensitivity of Dark Matter experiment is needed
in order to obtain relevant data concerning WIMP dark matter.

Here we present first results of the Heidelberg
Dark Matter Search (HDMS) experiment \cite{prophdms,hdms}, which took data over a period
of about 15 months in the Gran Sasso Underground Laboratory at LNGS in Italy.
After a description of the experimental setup, the
performance of the detectors is discussed. The last 49 d of data taking
are then analyzed in terms of WIMP-nucleon cross sections and a
comparison to other running dark matter experiments is made. 

In the following section we introduce the GENIUS-TF \cite{geniustf,la_geniustf}, 
a new experimental setup to probe the evidence region favoured by the
DAMA experiment \cite{damahere} and to test the prerequisites
neccessary to realize the Genius project \cite{genius,geniusbeyond}.

\section{The HDMS experiment}
\begin{figure}
\begin{center}
\includegraphics[width=.35\textwidth]{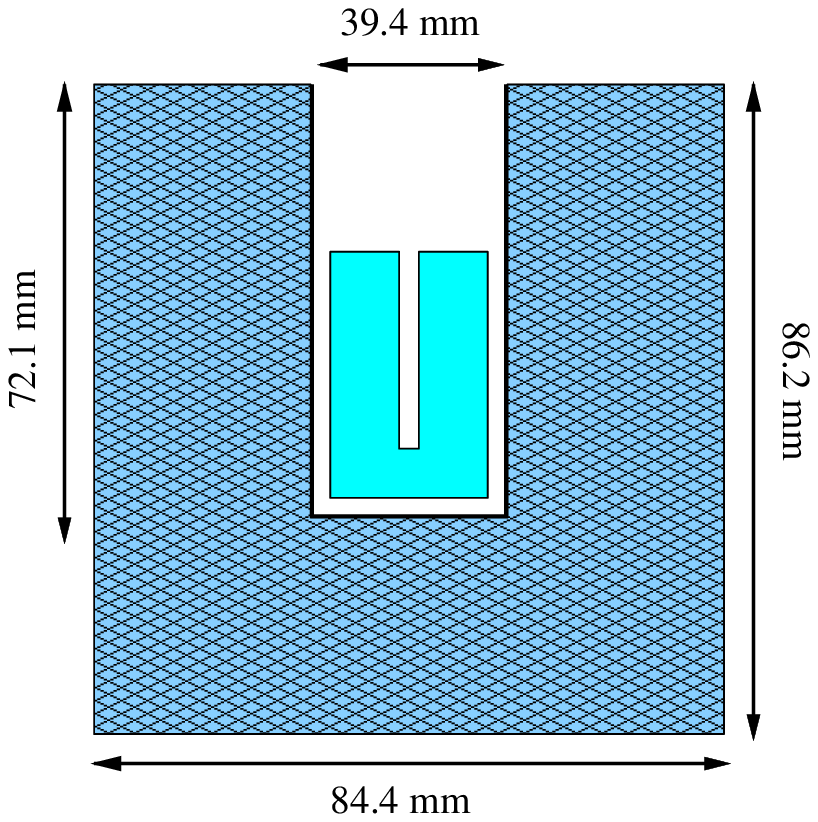}
\includegraphics[width=.496\textwidth]{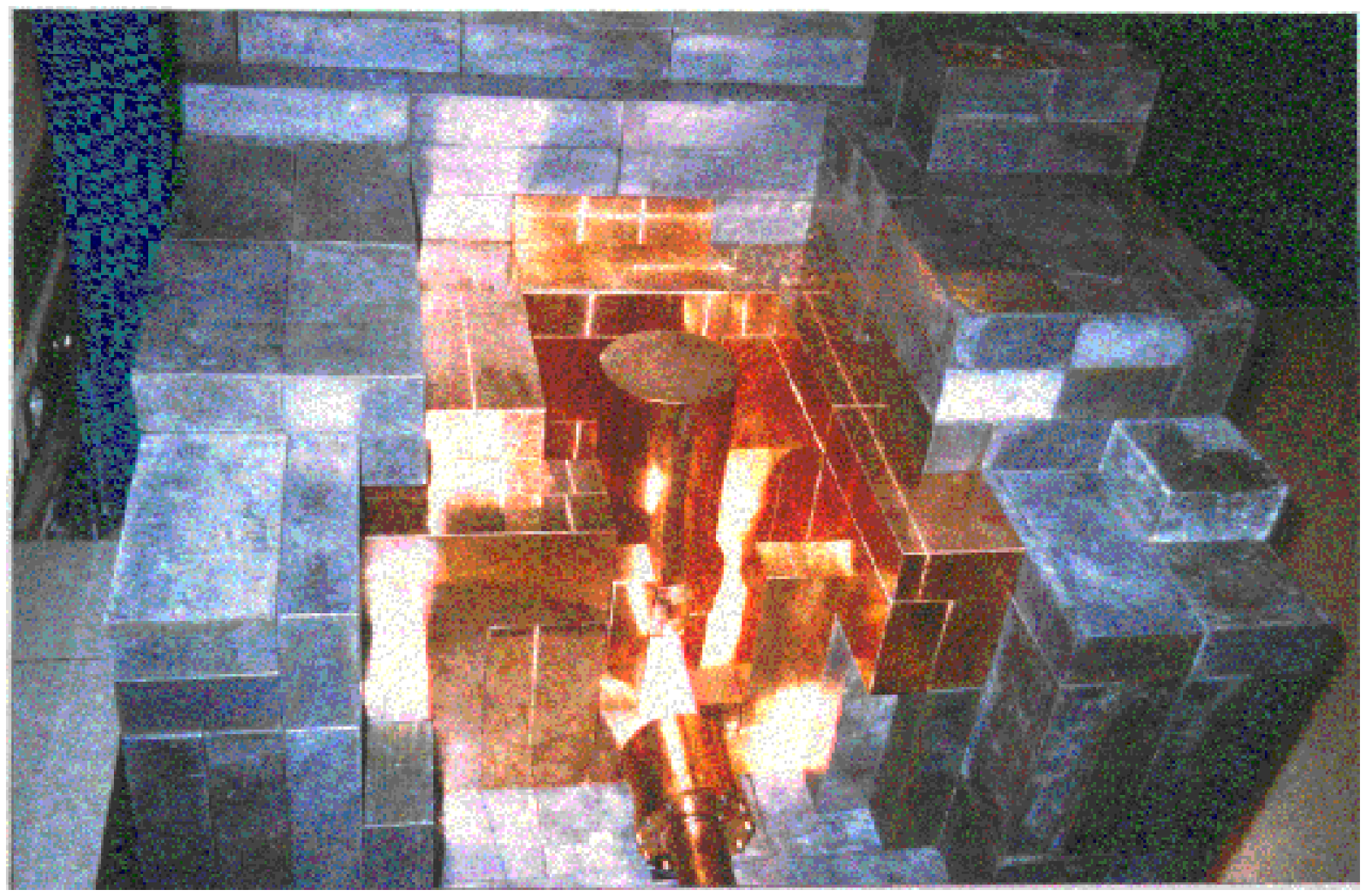}
\end{center}
\caption[]{Left: Schematic view of the HDMS experiment. A
  small Ge crystal is surrounded by a well type Ge-crystal,
  the anti-coincidence between them is used to suppress background
  created by external photons. Right: The HDMS detector during its
  installation at LNGS.} 
\label{detindet}
\end{figure}

HDMS operates two ionization HPGe detectors in a unique configuration \cite{hdms}.
A small, p-type Ge crystal is surrounded by a well-type Ge crystal,
both being mounted into a common cryostat system (see
Figure~\ref{detindet} for a schematic view). To shield leakage
currents on the surfaces, a 1\,mm thin vespel insulator is placed
between them. Two effects are expected to reduce the background of the
inner target detector with respect to our best measurements with the
Heidelberg-Moscow experiment \cite{WIMPS}. First, the anticoincidence
between the two detectors acts as an effective suppression of multiple
scattered photons. Second, we know that the main radioactive
background of Ge detectors comes from materials situated in the immediate
vicinity of the crystals. In the case of HDMS the inner detector is
surrounded (apart from the thin isolation) by a second Ge crystal -
one of the radio-purest known materials.

In order to house both Ge crystals and to establish the two high voltage and two
signal contacts, a special design of the copper crystal holder system was
required. 
The cryostat system was built in Heidelberg and made of low
radioactivity copper, all surfaces being electro-polished.
The FETs are placed 20 cm away from the crystals, their effect on the
background is minimized by a small solid angle for viewing the
crystals and by 10 cm of copper shield.

\subsection{Detector Performance at LNGS}

The HDMS prototype was installed at LNGS in March 1998.
Figure \ref{detindet} shows the detector in its open shield. The
inner shield is made of 10 cm of electrolytic copper, the outer
one of 20 cm of Boliden lead.  The whole setup is enclosed in an air
tight steel box and flushed with gaseous nitrogen in order to suppress
radon diffusion from the environment. Finally a 15 cm thick borated
polyethylene shield surrounds the steel box in order to minimize the
influence of neutrons from the natural radioactivity and muon produced
neutrons in the Gran Sasso rock.

The prototype detector successfully
took data over a period of about 15 month, until July 1999.
The individual runs were about 0.9 d long. Each day the experiment
was checked and parameters like leakage current of the detectors,
nitrogen flux, overall trigger rate and count rate of each detector
were checked. The experiment was calibrated weekly with a $^{133}$Ba
and a $^{152}$Eu-$^{228}$Th source. 
The energy resolution of both detectors (1.2\,keV at 300\,keV inner
detector and 3.2\,keV at 300\,keV outer detector) were stable as a function of time.
The zero energy resolution is 0.94\,keV for the inner detector and 3.3
keV for the outer one.

The energy thresholds are 2.0\,keV and 7.5\,keV for the inner and outer
detector, respectively.

\begin{figure}  
\begin{center}
\includegraphics[width=.496\textwidth]{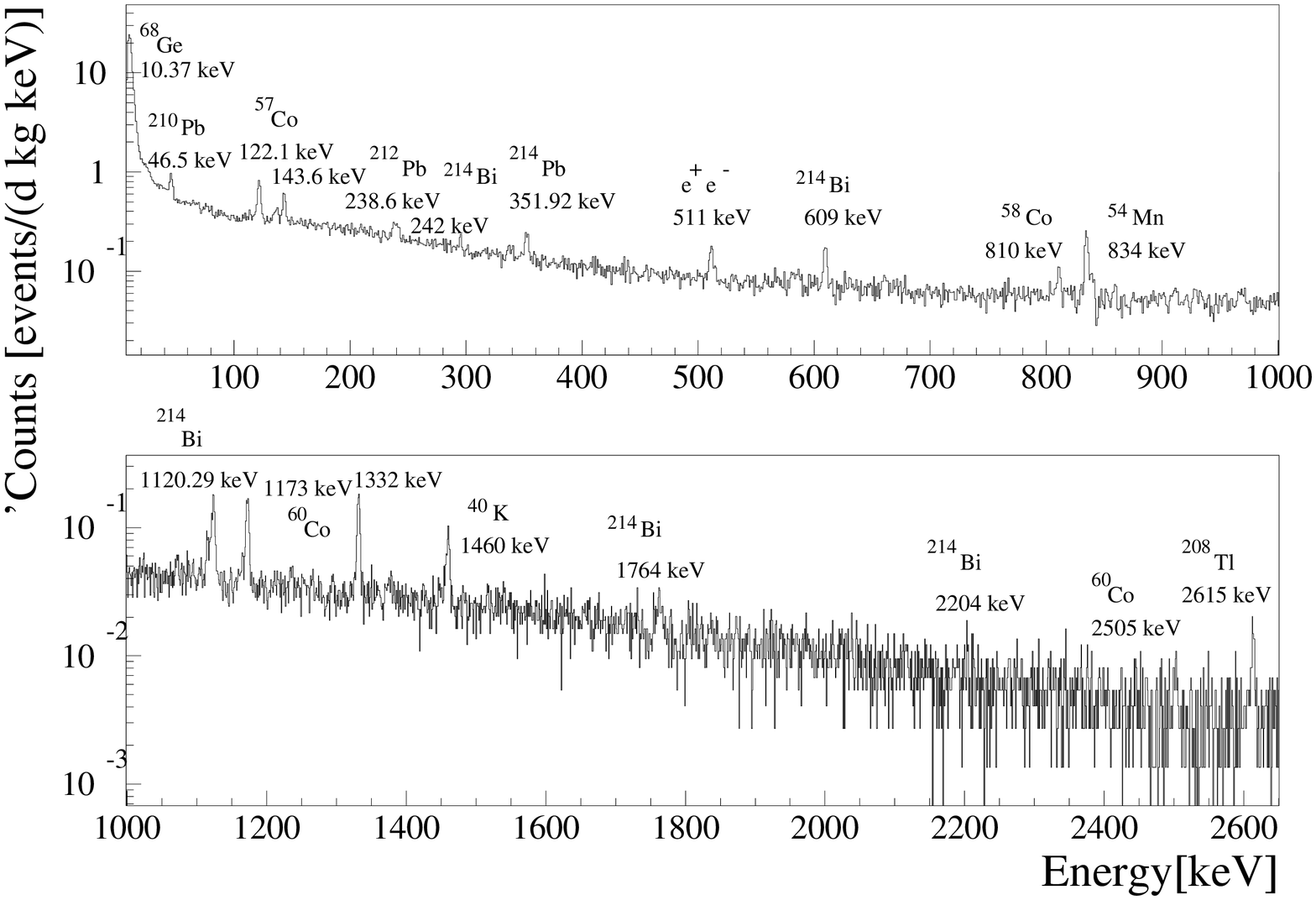}
\includegraphics[width=.496\textwidth]{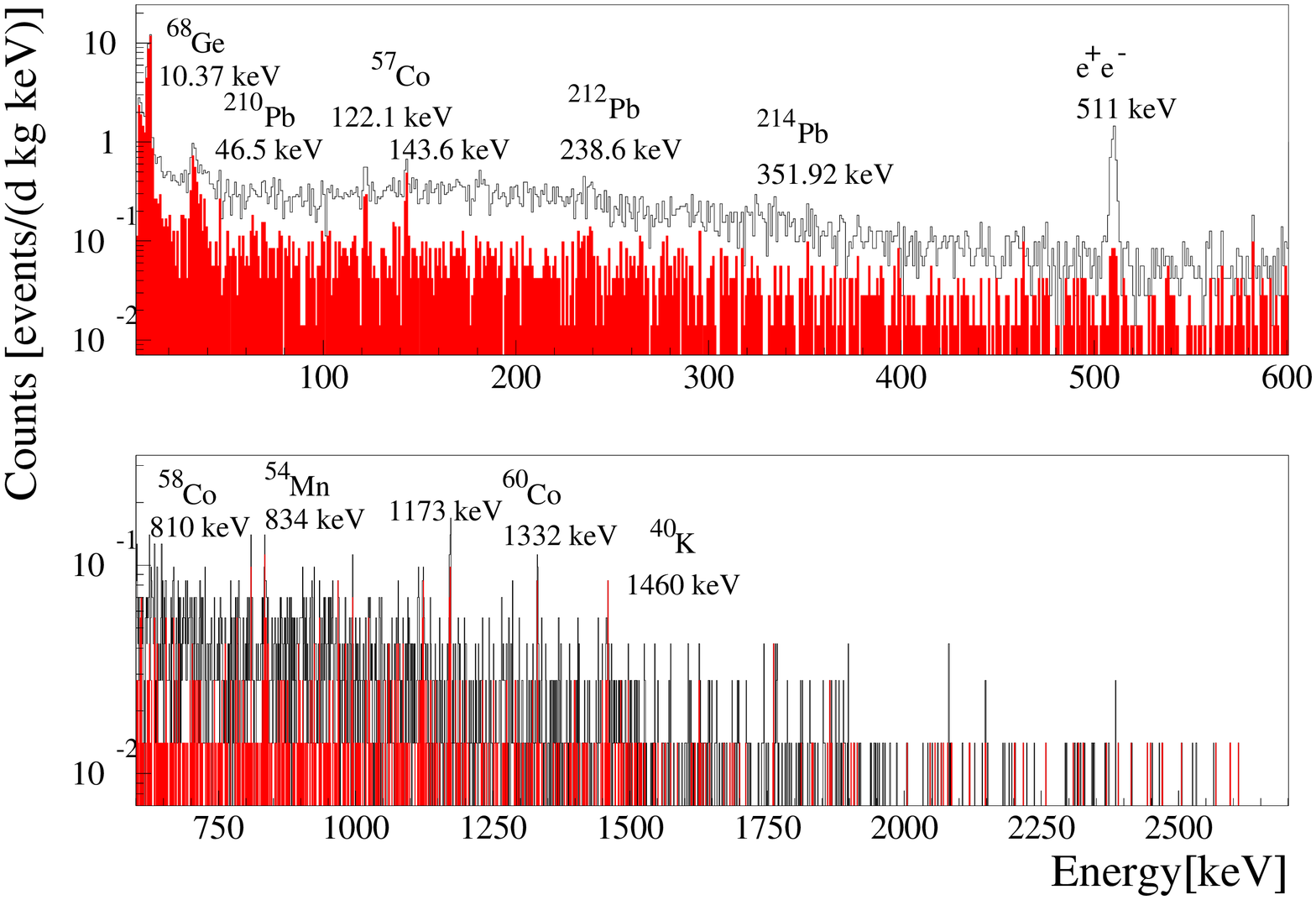}
\end{center}
\caption[]{Left: Sum spectrum of the outer detector after a
  measuring time of 363 days. The most prominent lines are labeled.
  Right: Sum spectrum of the inner detector after a
  measuring time of 363 days. The most prominent lines are
  labeled. The filled histogram is the spectrum after the
  anti-coincidence with the outer detector.
}
\label{sum-outer} 
\end{figure}

Due to the very special detector design, we see a cross-talk
between the two detectors. The observed correlation is linear and can
be corrected for off-line \cite{dm98}.
After correction for the cross talk and recalibration to standard
calibration values, the spectra of the daily runs were summed. Figure
\ref{sum-outer} shows the sum spectra for the
outer and inner detector, respectively (the most important identified 
lines are labeled).

In the outer detector lines of some cosmogenic and anthropogenic
isotopes, the U/Th  natural decay chains and $^{40}$K are clearly
identified.
The statistics in the inner detector is not as
good, however the X-ray at 10.37\,keV resulting from the decay of
$^{68}$Ge, some other cosmogenic isotopes, $^{210}$Pb and  $^{40}$K
can be seen.
The region below 10\,keV is dominated by the X-rays from cosmogenic
radio nuclides. In addition a
structure centered at 32\,keV is identified. Its
origin has recently been understood to be due to an artefact of the
special detector configuration.

\begin{figure}[t!]
\rotatebox{-90}{
\includegraphics[width=.675\textwidth]{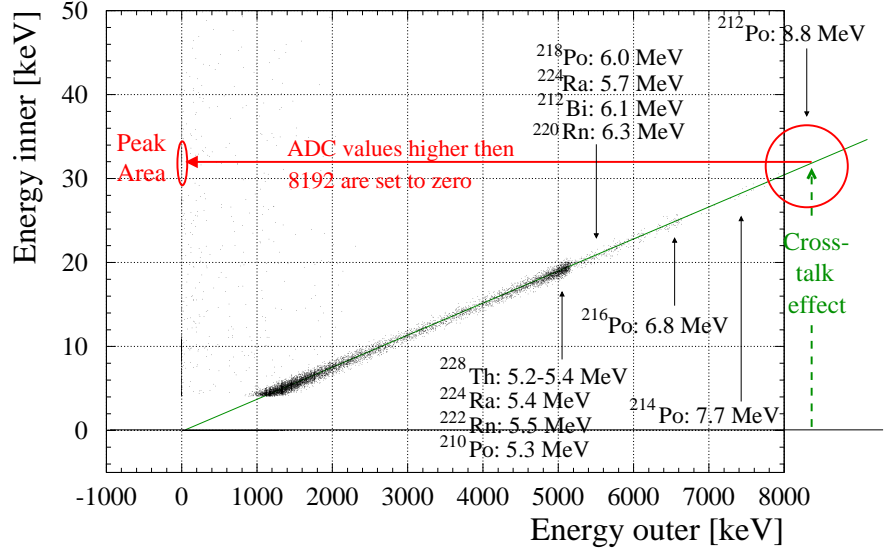}}
\caption{\footnotesize Origin of the structure observed at 32~keV in the inner
  detector: Shown is a scatter plot of all events with energies less
  then 50~keV in the inner detector. Each dot corresponds to one
  event. The position on the y-axis corresponds to the recorded energy
  in the inner detector, while the position on the x-axis
  displays the recorded energy in the outer detector. The linear
  dependence of the 
  pick-up in the inner detector due to the energy deposition in the
  outer detector is nicely seen. Clearly visible are regions
  dominated by $\alpha$-particles due to the decay 
  chains of $^{238}$U and $^{232}$Th. The highest energetic
  $\alpha$-particles resulting from these decay chains stem from
  $^{212}$Po with 8.8~MeV. Due to the limited dynamic range of the
  ADCs an event with such high energy can not be stored in the memory
  and is thus set to zero. Like this the event only appears with its
  crosstalk signal of $\sim$32.5~keV in the inner detector, faking a 
  peak, which is not affected by the anticoincidence.
  \label{structure}
  }
\end{figure}

In Fig. \ref{structure}
a scatter plot of all events with energy depositions less then
50~keV in the inner detector is shown.
The linear dependence of the pick-up signal in the inner detector 
from the energy deposition in the outer detector is nicely seen. 
The largest fraction of events with energies above 3~MeV in the outer
detector do result from $\alpha$ contaminations. These stem from the
decay-chains of $^{238}$U and $^{232}$Th, most probably resulting from 
U/Th contaminations being present
in the soldering tin of the contacts for the outer detector. The
highest energetic 
$\alpha$-decays with 8.8~MeV result from the
decay of $^{212}$Po. However, for such high energies (above $\sim$ 8~MeV) the
dynamic range of the ADC does not allow for the recording of the
event. The ADC value is set to zero in such a case. The crosstalk seen
in the inner detector can be calulated once the slope of the pick-up 
is known. This has been measured to be $k_{io}=0.00375\pm 0.00004$.
Thus only a
pick-up event in the inner detector with energies around 
$0.00375\times 8800\sim 33.0$~keV is
recorded whereas no event seems to occur in the outer detector
(some energy loss occurs due to the
dead layer of the crystals, thus the slightly higher energy value).

After 363 days of pure measuring time the statistics in the inner
detector was high enough in order to estimate the background
reduction through the anti-coincidence with the outer detector.

\begin{figure}  
\begin{center}
\includegraphics[width=.6\textwidth]{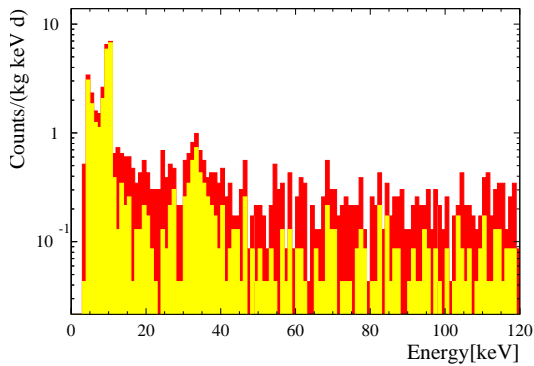}
\end{center}
\caption[]{Low-energy spectrum of the inner detector.
  The light shaded spectrum
  corresponds to the events using the anticoincidence, the dark shaded
  spectrum to all events.}
\label{low-inner} 
\end{figure}

Figure \ref{low-inner} shows the low-energy spectrum of the inner detector
before and after the anti-coincidence. 
The cosmogenic X-rays below 11\,keV are preserved, since in this case
the decays are occurring in the inner detector itself. Also a $^3$H
spectrum with endpoint at 18.6\,keV is presumably present.
If the anti-coincidence is evaluated in the energy region between
40\,keV and 100\,keV, the background reduction factor is 4.3. 
The counting rate after the anti-coincidence in this energy region is 
0.07\,events/(kg\,d\,keV), thus very close to the value measured in the 
Heidelberg-Moscow experiment with the enriched detector ANG2 \cite{WIMPS}.    
In the energy region between 11\,keV and 40\,keV the background index is 
with 0.2\,events/(kg\,d\,keV) a factor of 3 higher.

\subsection{Dark Matter Limits}

The evaluation for dark matter limits on the WIMP-nucleon cross section 
$\sigma_{\rm scalar}^{\rm W-N}$ follows the conservative 
assumption that the whole experimental spectrum consists of 
WIMP events.
Consequently, excess events from calculated 
WIMP spectra above the experimental spectrum in any energy region 
with a minimum width of the energy resolution of the detector 
are forbidden (to a given confidence limit). 

The parameters used in the calculation of expected WIMP spectra are 
summarized in \cite{WIMPS}. We use formulas given 
in the extensive review \cite{lewin} for a truncated 
Maxwell velocity distribution in an isothermal WIMP--halo 
model (truncation at the escape velocity, compare also \cite{freese}). 

After calculating the WIMP spectrum for a given WIMP mass, the scalar 
cross section is the only free parameter which is then used to fit the 
expected to the measured spectrum 
using a one-parameter maximum-likelihood fit algorithm. 

\begin{figure}
\begin{center} 
\includegraphics[width=0.6\textwidth]{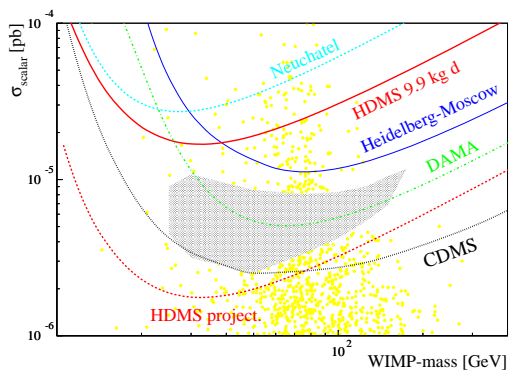}
\end{center}
\caption{
  WIMP-nucleon cross section limits as a function of the WIMP
  mass for spin-independent interactions. 
  The upper solid line corresponds to the limit set by the HDMS-prototype
  detector. 
  The plain curves correspond to the limits given by CDMS \cite{rick2000},
  DAMA \cite{damaexcl} and the Heidelberg-Moscow Experiment \cite{WIMPS}.
  The
  filled contour represents  the 2$\sigma$ evidence region of the DAMA
  experiment \cite{damahere}. The lower dashed line shows the
  expectation for the HDMS final setup assuming a threshold of 2~keV.   
  }
\label{dm_limits}
\end{figure}

To compute the limit for the HDMS inner detector we took only the last 
49 days of measurement. We omit the first 260\,days in order to
reduce the contaminations due to long-lived cosmogenically produced
materials. These have life times of typically $\sim$~200 days.
The energy threshold of the measurement was 2\,keV.
The resulting preliminary upper limit exclusion plot in the 
$\sigma_{\rm{scalar}}^{\rm{W-N}}$ versus M$_{\rm{WIMP}}$
plane is shown in Fig.~\ref{dm_limits}.

Already at this stage, the limit is competitive with our limit from
the Heidel\-berg-Moscow experiment. In the low mass regime for WIMPs the 
limit has been improved due to the low-energy threshold of 2~keV
reached in this setup. 

Also shown in the figure are limits from the Heidelberg-Moscow
experiment \cite{WIMPS},
limits from the DAMA experiment \cite{damaexcl} and the most recent
results in form of an exclusion curve from the  
CDMS experiment \cite{rick2000}. The filled contour represents  the
2$\sigma$ evidence region of the DAMA experiment \cite{damahere}.    

\subsection{Outlook for the HDMS Experiment}

The prototype detector of the HDMS experiment successfully took data 
at LNGS over a period of about 15 months. 
All of the background sources were identified. The background reduction
factor in the inner detector through anticoincidence is about 4. It is 
less then previously expected \cite{hdms}, due to the smaller
diameter of the veto detector than originally planned.
Nevertheless, the background in the low-energy region of the inner
detector (with exception of the region still dominated by cosmogenic
activities) is already comparable to the most sensitive dark matter
search experiments.

For the final experimental setup, important changes were made.
\begin{itemize}
\item The crystal holder was replaced by a holder made of ultra low
level copper.
\item The soldering of the contacts was avoided, thus no soldering tin 
  was used in the new setup.
\item The inner crystal made of natural Germanium in the described
  prototype was replaced by an enriched $^{73}$Ge crystal (enrichment
  86\,\%). In this 
  way, the $^{70}$Ge isotope (which is the mother isotope for
  $^{68}$Ge production) is
  strongly de-enriched (the abundance in natural Germanium is 7.8\%).
\end{itemize}

After a period of test measurements in the low-level laboratory in
Heidelberg, the full scale experiment was installed at LNGS in the 
in August 2000. The projected final sensitivity of the detector can be 
seen in Fig.~\ref{dm_limits}.

\section{The GENIUS experiment}
In order to achieve a dramatic step forward regarding background reduction, 
a new experimental technique is
needed.

Lately there have been two promising approaches to reach this goal:
\begin{itemize}
\item Application of standard detection techniques while removing all 
        dangerous contaminations from the direct vicinity of the detectors.
\item Cryo-detectors have been developed which are able to detect two signals
        of the WIMP-nuclear recoil phonons and ionization (e.g. the
        CDMS experiment \cite{cdmshere}), or phonons
        and scintillation (e.g. the CRESST phase 2 
        experiment \cite{seidel}) simultaneously. 
        This enables a very effective discrimination between nuclear recoils
        and electromagnetic interactions.
\end{itemize}

The GENIUS (GErmanium in liquid NItrogen Underground Setup) proposal
uses the first concept \cite{genius,geniusbeyond}.

\begin{figure}
\begin{center}
\includegraphics[width=.496\textwidth]{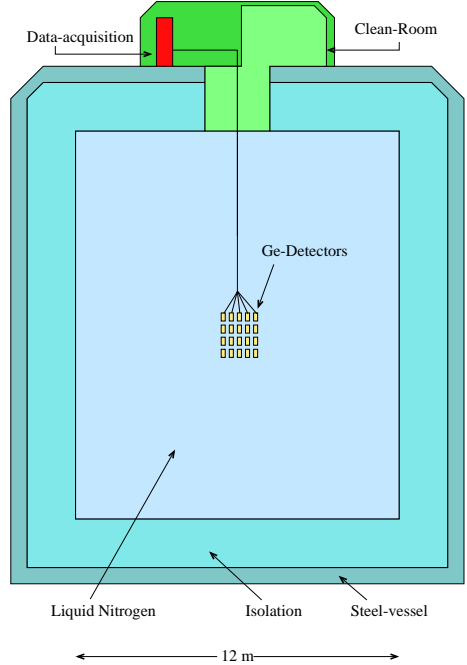} 
\end{center}
\caption[]{\label{genius_scheme}Schematic view of the GENIUS project. An array 
of 100~kg of natural HPGe detectors for the WIMP dark matter search (first step) 
or between 0.1 and 10 tons of
enriched $^{76}$Ge for the double beta decay search (final setup) is hanging 
on a support
structure in the middle of the tank immersed in liquid nitrogen. The size of
the nitrogen shield would be 12 meters in diameter at least. On top of the tank
a special low-level clean room and the room for the electronics and data 
acquisition will be placed. }
\end{figure}

\subsection{The concept of the GENIUS experiment}
The GENIUS project \cite{genius,geniusbeyond} 
is based on the idea to operate 'naked' HPGe crystals
directly in liquid nitrogen \cite{heusser2} and to remove all dangerous
contaminations from the direct vicinity of the crystals. That Ge
detectors really work in liquid nitrogen has been shown in
\cite{GENIUS,NIM}. Using a 
sufficiently large tank of liquid nitrogen, the latter can act
simultaneaously as cooling medium and shield against external
activities, since it is very clean
with respect to radiopurity due to its production history (fractional
distillation). The conceptual design of the experiment is
depicted in figure \ref{genius_scheme}.

The proposed detection technique is based on ionization in HPGe detectors. 
The crystals would be of p-type.
p-type detectors have the advantage that the outer contact 
is n$^+$ and the surface dead layer therefore several hundred micrometers.
This effectively prevents the detection of $\alpha$- and $\beta$ particles
which would otherwise dominantely contribute to the background.
The ideal working temperature of the p-type detectors is 77 K.
The cooling of the HPGe crystals is very efficient since the detectors are 
in direct thermal contact with the cooling medium liquid nitrogen.

\begin{figure}
\begin{center}
\includegraphics[width=.8\textwidth]{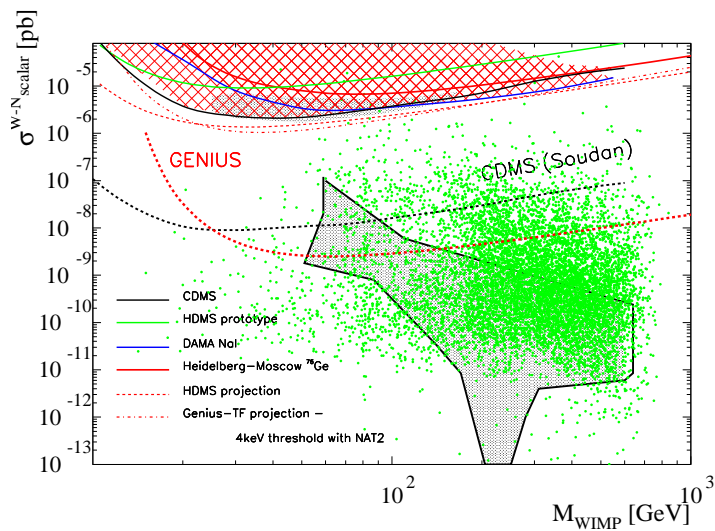}
\end{center}
\caption[]{\label{exclusion_intro} Exclusion plot of the scalar 
WIMP-nucleon elastic scattering cross 
section as a function of the WIMP mass. Plotted are excluded areas from
the presently most sensitive direct detection experiments (hatched area,
DAMA \cite{damaexcl}, CDMS \cite{cdmshere}, Heidelberg-Moscow \cite{WIMPS}, HDMS prototype\cite{hdms}) 
and some projections for experiments running or being presently under 
construction (HDMS, GENIUS-TF \cite{la_geniustf}). 
The extrapolated sensitivities of future
experiments (GENIUS \cite{geniusbeyond}, CDMS at Soudan \cite{cdmshere}) 
are also shown. The scatter plot 
corresponds to predictions from theoretical considerations of
the MSSM \cite{klanew}. 
The small shaded area represents the 2$\sigma$ evidence region from the
DAMA experiment \cite{damahere}. 
The large shaded area corresponds to calculations in the 
mSUGRA-inspired framework of the MSSM, with universality relations for the
parameters at GUT scale \cite{ellis} (Figure taken from \cite{klanew}).}
\end{figure}

It has been shown that according to Monte Carlo simulations with this
approach a 
reduction of background by three to four orders of magnitudes can be achieved
\cite{genius,geniusbeyond,GENIUS,NIM}. The final reachable
background index is estimated to be around $\sim 10^{-2}$
counts/(kg~keV~y) in the low-energy region below 100~keV relevant for
WIMP Dark Matter search. The sensitivity reachable with this background 
for neutralinos as dark matter can be seen in Fig. \ref{exclusion_intro}.

\subsection{Tritium production in HPGe at sea level}

As evident from previous considerations of the expected background
\cite{geniusbeyond,NIM}, great care has to be taken about the
cosmogenic isotopes produced inside the HPGe crystals at sea level. 
Without additional shield against the hard component of cosmic rays during
a fabrication time of ten days many isotopes are
produced which significantly reduce the sensitivity of GENIUS as a dark 
matter detector. Especially the production of $^{68}$Ge from the isotope
$^{70}$Ge affects the sensitivity by
increasing the energy threshold of the detector to 12~keV.
In the main reaction leading to $^{68}$Ge enhancement also
tritium is produced through the process
$^{70}$Ge(n,t)$^{68}$Ge. Tritium has a half life of 
12.35 years and can thus not be deactivated within a reasonable time.
$^3$H is a $\beta$ emitter with a Q-value of 18.6~keV. 

The cosmogenic production rate of $^3$H in natural germanium 
has been estimated  through simulations in \cite{juan,avignone} using the 
cosmic neutron fluxes cited in \cite{lal,hess}. 
For natural germanium it is estimated 
to be less than $\sim$ 200 atoms per day and kg material. 
Using this upper limit for tritium production at sea level
with an overall fabrication time of ten days this would
mean a tritium abundance of $\sim$ 2000 atoms per kilogram material.
With the half life of 12.3 years this results in a decay rate of
$\sim$3.6 $\mu$Bq/kg equivalent to $\sim$113 decays per year (this is in
very good agreement with the result in \cite{zdesenko}).
Even assuming an energy threshold of 12~keV and taking into account the 
spectral shape of tritium decay this yields
an event rate of approximately 2 counts/(kg~keV~y) in the energy
region between 
12~keV and 19~keV, which is by two to three orders of magnitudes above the 
allowed count rate.

This consideration drastically shows the importance of proper planning
of the crystal production and transportation.
To avoid major problems with cosmogenic isotopes it is
therefore essential to minimize the exposure of the crystals to cosmic rays at
sea level.

If it is assumed that during the zone refining process the
germanium material is sufficiently shielded against the cosmic
radiation, the  
unshielded time would with $\sim$ 3.5 days be by two orders of
magnitudes too long. 
This exposure at sea level would result in approximately 0.7
counts/(kg~keV~y).
In addition it has to be taken into account that
the crystal might have to go through several production steps more
then once 
thus increasing the exposure time by another 16 hours per 
additional cycle.

It is therefore required to additionally shield the detector material
during production and transportation using approximately 2m of heavy concrete.
Heavy concrete can be produced with a density up to 5.9~g/cm$^3$. Thus an
additional concrete shield of 1~m could act as a shield of roughly 5 mwe.
This reduces the hard nucleonic component mainly responsible for the 
cosmogenic isotope production by one to two orders of magnitudes 
\cite{heusser2}. A further increase of shielding strength does not seem to be
reasonable since the cosmogenic production through the cosmic fast muons 
which is by approximately two orders of magnitudes less than through the 
hadronic component can not be shielded whatsoever.

To make a first approximation of the tritium abundance in the crystals
after production and transportation, it is assumed that a shield of 5 mwe 
can be provided
during both fabrication and transport, resulting in a reduction of tritium 
production by a factor of $\sim$ 30 (see figures 2 and 3 in \cite{heusser2}).
The time interval relevant for tritium 
enhancement starts directly after the zone refining process, since in this
step most of the contamination is being removed from the germanium material.
Thus for the fabrication (in the ideal case) 78 hours are needed.
Without considering transport, this results in approximately 20 tritium atoms
per kg detector material. Taking into consideration also a transportation
time of one week (shipping), the amount of produced tritium atoms 
increases to $\sim$70 atoms per kg.

The expected decay rate is $\sim$1.1 per year without and $\sim$3.9 per year 
with transportation considered.
 If an energy threshold
of 12~keV is assumed for the experiment due to the decay of $^{68}$Ge,
the events resulting from tritium decay below 12~keV can be neglected.
Due to the spectral shape of tritium, every $\sim$10th decay deposits 
more than 12~keV of energy in the detector.
In the energy interval between 12~keV and 19~keV thus 0.11 events per year 
and 0.39 events per year
are expected from tritium without and with transportation 
considered, respectively.
The final background sensitivity would therefore be 
$\sim$1.6$\times10^{-2}$ counts/(kg~keV~y)
without additional transportation and $\sim$5.6$\times10^{-2}$
counts/(kg~keV~y) with a 
week of transport from the fabrication site to the site of the experiment.
This background level almost corresponds to the curve shown in
Fig. \ref{exclusion_intro}. 

Note that the consideration made here is a very crude approximation. It
is, however, possible to say that tritium will definitely limit the 
sensitivity of GENIUS as a dark matter detector if the germanium crystals
are not produced directly underground. Thus it should be seriously considered
to produce the detectors underground, directly at the experimental site.

\section{The GENIUS-TF}

It has been shown in the BARGEIN proposal \cite{bargein} that with a setup
using a conventional shield, a sensitivity can be reached which allows for a
test of the DAMA evidence region within a short time period.
However, with the BARGEIN setup this test could only be done looking for the 
expected signal of WIMP dark matter in HPGe detectors: the WIMP-nucleus recoil
spectrum. 
Thus the BARGEIN setup can only verify the CDMS result of (almost)
excluding the possibility of WIMP dark matter as favored by the DAMA
data \cite{rick2000}.  
If the active mass of the detector can be increased to approximately
40 kg, as proposed for the GENIUS Test-Facility \cite{geniustf,la_geniustf},
and the background index will be
maintained, also the expected signature of WIMP dark matter in form of the
annual modulation signal could be tested within a reasonable time window
\cite{geniustf,la_geniustf}. 

The primary goal of  GENIUS-TF is to demonstrate the feasibility of
the GENIUS project.

With the GENIUS TF it is planned to test the
following points:

Material selections have to be performed for various experimental
components like polyethylene and contacting wires, and their purities
have to be tested down to 1\,event/(kg\,y\,keV). A crystal support system, 
made of low-radioactivity polyethylene has to be developed and
designed such that it 
can be extended in order to house up to 40 crystals (100\,kg) and more
and later implemented into the GENIUS setup. 

Furthermore a
new, modular data acquisition system and electronics, capable of
taking data from up to 300 and more detectors simultaneously has have
to be developed and tested. 

From the low-energy spectrum valuable experiences can also be gained about
the cosmogenic activation of the HPGe-crystals, since the exposure history of 
the crystals is monitored in detail during manufacturing and
transport. Especially for the $^{3}$H contamination it is of utmost
importance to have a basic understanding of its production rate for
the GENIUS project. 

Finally it has  to be confirmed that `naked' Ge detectors work reliably in 
liquid nitrogen over a longer period of time. In the former studies
the HPGe crystals have been operated reliably over typical time periods 
of weeks; however, for an experiment of the scale of the GENIUS
project, it has to be shown that this is also possible over time
scales of years.

Besides above issues, GENIUS-TF will have a physics program of its
own, as discussed below.

\subsection{The Test Facility}
The concept of the GENIUS proposal has the great advantage that no 
individual cryostat system is needed.
Instead the HPGe crystals are surrounded by liquid nitrogen of much
higher radiopurity which in addition provides ideal cooling and
shielding against external radiation. This opens the new research
potentials for the Genius project \cite{genius,geniusbeyond,GENIUS}.

It is proposed to install a setup with up to fourteen detectors 
on a small scale \cite{geniustf,la_geniustf} in order to be sensitive in the range
of the DAMA 
result \cite{damahere} on a short time scale and to prove the long term
stability of the new detector concept and some other important aspects
for the realization of the GENIUS project discussed above.

\begin{figure}
\begin{center}
\includegraphics[width=.75\textwidth]{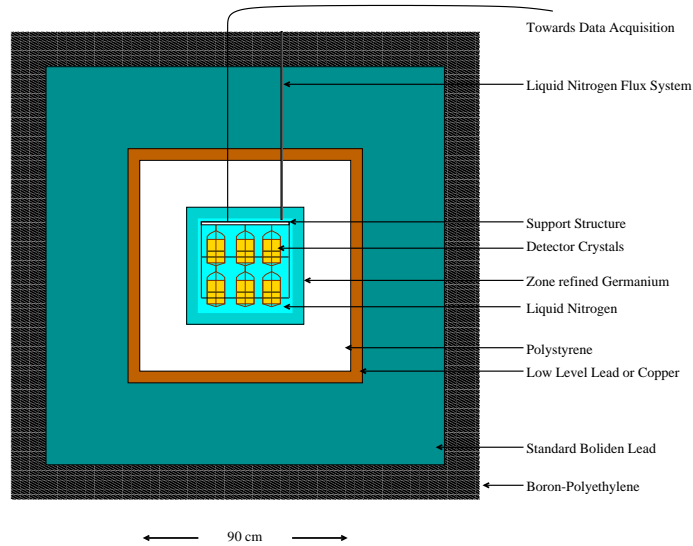}
\end{center}
\caption[]{\label{geniustf_scheme} Conceptual design of the GENIUS-TF. Up
to 14 detectors will be housed in the inner detector chamber, filled with
liquid nitrogen. As a first shield 5 cm of zone-refined Germanium will be used.
Behind the 20 cm of polystyrene isolation another 35 cm of low-level lead
and copper and a 15 cm borated polyethylene shield will complete the setup.}
\end{figure}

The design is shown in figure \ref{geniustf_scheme}. 
It is based on a dewar made from low activity polystyrene
and on a shield of zone-refined germanium bricks inside the dewar and
low-activity lead outside the dewar. A layer of boron-loaded
polyethylene plates for suppression of neutron-induced background
completes the shield.

330 kg of zone-refined
high-purity Germanium bricks would serve as the inner 
layer to shield the 'naked' HPGe detector against the less radio-pure
polystyrene.  Also the first 10cm layer outside the
polystyrene-dewar needs to be of extreme radiopurity. The same type of 
copper as
installed in the Heidelberg-Moscow experiment, and/or some complementary
low-level lead could be used.
To shield the external $\gamma$ rays (natural radioactivity from the
surroundings) an overall lead layer of approximately 35 cm is needed. 

Using this concept an inner detector chamber of
40~cm$\times$40~cm$\times$40~cm 
would be sufficient to house up to seven HPGe-detectors in one layer or
14 detectors in two layers. This will 
allow for the development and test of a holder system for the same
amount of crystals.

The overall dimension of the experiment will be 
1.9~m$\times$1.9~m$\times$1.9~m (without the boron-loaded
polyethylene) thus fitting in one of the buildings 
of the Heidelberg-Moscow experiment which is used momentarily for
material measurements. 

The background considerations and simulations discussed in
\cite{la_geniustf,bargein} suggest  that a reduction of the background by
a factor 
of $\sim$5 with respect to the Heidelberg-Moscow-Experiment can be
attained with the proposed setup.

Assuming a final target mass of 40~kg, an energy threshold of 12~keV and
a background index of 4 counts/(kg keV y)
corresponding to $\sim$ 0.01 counts/(kg keV d) in the energy region between 
12~keV and 100~keV the GENIUS-TF would need a significance of
190~kg~y to see the claimed DAMA annual modulation with 95\% probability
and 90\%C.L. (see \cite{cebrian}). This corresponds to an overall measuring 
time of approximately five years which would correspond to the life time of
this experiment.

However, the new detectors will have
an energy threshold of 0.5~keV (four detectors have already been
produced by the end of February 2001) thus allowing for the use of the
experimental 
spectrum in the energy range between the threshold and the X-ray peaks seen
from the cosmogenically produced isotopes. This could significantly improve
the sensitivity of the GENIUS-TF on the annual modulation effect.

\subsection{Installation of the GENIUS-TF and time schedule}
With the dimensions for the inner detector chamber given above, the materials 
shown in table \ref{materialien} will be needed for the installation of 
the GENIUS-TF.

The Heidelberg-Moscow-Collaboration possesses $\sim$ 330 kg of zone refined
Germanium bricks, $\sim$ 10 tons of Boliden-lead and $\sim$ 500 kg
ancient low-level lead. These materials can be installed in the
GENIUS-TF without additional costs. 

\begin{table}
\begin{centering}
\begin{tabular}{lcr}
\hline
Material             & Amount\\
\hline
14 HPGe-detectors     & $\sim$ 40 kg\\
\hline
Electronics          &  ---          \\
\hline
Germanium bricks\footnotemark[2]      & $\sim$ 330 kg\\
Polystyrene box          & $\sim$ 40 kg  \\
Low-Level Lead (LC2 or ancient) or& 3076 kg \\
Low-Level Copper  & $\sim$ 7 t       \\
Boliden Lead bricks  &  $\sim$ 30 t    \\
\hline
Nitrogen filling and & 3-4 100 liter low\\
cleaning device      & level dewars \\
\hline
\hline
\end{tabular}
\caption{\label{materialien} Amount of materials 
needed for the installation of the GENIUS-TF.}
\end{centering}
\end{table}

The BOREXINO collaboration is running a liquid nitrogen filtering device
in the Gran Sasso underground laboratory. The capacity of this machine is 
by far not used by the needs of the BOREXINO collaboration. Thus this device
could also serve as the low-level nitrogen support for the GENIUS-TF. 
Once a week two hundred liters of low-level nitrogen dewars could be
filled from the 
filtering device and stored for deactivation of $^{222}$Rn for one week. This
amount of liquid nitrogen would be enough for approximately one week.

\footnotetext[2]{Property of the Heidelberg-Moscow-Collaboration}

The development of the liquid nitrogen cleaning
and filling device will be started soon in collaboration with
the BOREXINO experiment. The construction of the setup
can be started immediately since no additional space in the Gran
Sasso Underground Laboratory is required. 
The data acquisition system of the HDMS experiment can be used to
obtain first data with two detectors which are already housed in the 
Gran Sasso Underground Laboratory. The first results can be
expected in the end of the year 2002 already.

\section{Conclusions}
We presented first results of a 15 month measuring period with the
prototype HDMS detector. The obtained sensitivity is already now
comparable to the most sensitive dark matter search experiments. The
final setup has been installed in August 2000 in the Gran Sasso
Underground Laboratory. Several improvements have been achieved for
the final detector: The crystal holder system was replaced with a
low-level copper holder system and soldering of the contacts was avoided.
Furthermore the inner detector consists of enriched
$^{73}$Ge, thus strongly suppressing the $^{68}$Ge contamination
with respect to the prototype detector, substantially
lowering the energy threshold. The expected sensitivity of the HDMS
experiment will allow to test by exclusion plot the DAMA evidence
region from the annual modulation signature \cite{damahere}.

We proposed the GENIUS-TF, a new experimental setup which has been
approved. We showed, using detailed
Monte Carlo simulations that the GENIUS-TF could reach a background of 
$\sim$ 4\,counts/(kg\,keV\,y) in the energy region between 
12\,keV and 100\,keV. 
Thus it could for the first 
time probe the DAMA evidence region using both, the WIMP-nuclear
recoil signal and the annual modulation signature. The GENIUS-TF is planned
as a test setup for the GENIUS project and will be installed in 2001.

%

\end{document}